\documentclass[twocolumn,showpacs,preprintnumbers]{revtex4-1}
\usepackage{graphicx}

\newcommand{\ie}{{\it i.e.}}
\newcommand{\eg}{{\it e.g.}}

\begin{document}

\title{Instability of Bose-Einstein condensates in tilted lattices with time-periodical modulation}
\author{Ning-Ju Hui, Xiao-Qiang Xu, and You-Quan Li}
\affiliation{Zhejiang Institute of Modern Physics and Department of
Physics, Zhejiang University, Hangzhou 310027, People's Republic of
China}

\begin{abstract}
We study the dynamical stability of Bose-Einstein condensates in an optical lattice with a time-periodic modulation potential and a constant acceleration force simultaneously.
We derive the explicit expressions of quasienergies and obtain
the stability diagrams in the parameter space of the interaction strength and the modulation amplitude.
The ratio of the acceleration force to the modulation frequency characterizes two cases: integer and non-integer resonances.
For integer resonances, the critical interaction strength $g_{\mathrm{c}}$ shows an alternate behavior where the completely unstable regions correspond to the negative effective tunneling strength.
Among non-integer resonances, we observe that $g_{\mathrm{c}}$ peaks are centered around half-integer resonances for which the completely unstable regions disappear, accompanied with a whole displacement of $g_{\mathrm{c}}$. Compared with integer and half-integer resonances,
the crossovers between them show no explicit dependence of $g_{\mathrm{c}}$ on the modulation amplitude.
\end{abstract}

 \received{\today } \maketitle

\section{Introduction}

Since Bose-Einstein condensation (BEC) of dilute gases of alkali
atoms was realized in 1995~\cite{BECs-realize1, BECs realize2, BECs
realize3}, much attention has been paid to the ultracold atomic
systems in various configurations, such as optical lattices which
are formed by counter-propagating laser beams. Due to the
flexibility of the system, the famous Mott-insulator-Superfluid
transition of BECs in optical lattices has been observed and
studied~\cite{Mott-insulator, periodic the3}. Dynamically, there are
also many interesting phenomena, such as the resonant
tunneling~\cite{resonant tunneling}. In the limit of vanishing
particle interactions, $\eg$, in dilute bosonic gases, the system
may exhibit Bloch oscillation~\cite{tilted 1 and B-o, Bo 3}, similar
to the behavior of single electron in the crystalline field of
solids. Other than the energetic instability~\cite{instability the2}
induced by the non-zero temperature, the nonlinear scattering
process may also cause the dynamical instability~\cite{dynamical
insta 2} damping the oscillation behavior~\cite{tilted exp2}.

Dynamical stability, which describes whether the system remains
stable or not during its time evolution, is our concern in this
paper. We are interested in the dynamical response of BECs in
modulated optical lattices. We notice that dynamical stability of
BECs system in an optical lattice modulated by a time-periodic
potential has been investigated in both experiment~\cite{periodic
exp6} and theory~\cite{periodic the5}. Also, dynamical stability of
the condensates in tilted optical lattices considering the effect
of a constant acceleration force has attracted many people's
attention~\cite{tilted the2, tilted the3}. Recently BECs in tilted
time-periodically modulated lattices formed by both a time-periodic
modulation potential and a constant acceleration force was realized
experimentally~\cite{potential exp2}. We focus on the dynamical
stability of the above system to which few attention has been paid
before, and aim to find out the influence of the interparticle
interactions.

The structure of this paper is organized as follows. In
Sec.~\ref{sec:model} we describe our modulated optical lattice
system with modified Bose-Hubbard Hamiltonian~\cite{potential the1,
potential the3}. In mean-field approximation we derive the
time-evolution equations and obtain the explicit solutions. In order
to determine the dynamical stability of the system, we adopt an
ansatz of the time-evolution operator~\cite{periodic the5, tilted the2}. Upon Floquet theorem, the
quasienergy analysis, which is strongly related to the dynamical
stability, is given in Sec.~\ref{sec:quasienergy}. Based on the ratio of the acceleration force to the modulation frequency, two specific
cases, integer and non-integer resonances, are considered. We also
discuss the main features of dynamical stability for both two cases
in Sec.~\ref{sec:dynamical stability}. A summary with brief
discussion is given in Sec.~\ref{sec:summary}.

\section{MODEL AND METHOD}\label{sec:model}

We consider ultracold bosonic gases on an one-dimensional optical
lattice modulated by both a time-periodic potential and a constant
external acceleration force simultaneously. To model such a system,
we need to include the external modulation terms in the typical
Bose-Hubbard Hamiltonian (BHH), \ie,
\begin{eqnarray}
\label{eq:Ha}
 \hat{H}&=&
 -J\sum_{<m,n>}(\hat{a}^{\dag}_m\hat{a}_n+H.c.)+\frac{U}{2}\sum_m\hat{n}_m(\hat{n}_m-1)\nonumber\\
 &&+[K\cos(\omega t)+d]\sum_mm\hat{n}_m.
\end{eqnarray}
Here $\hat{a}_m~(\hat{a}^{\dag}_m)$ annihilates (creates) a boson on
the $m$th lattice site, and $\hat{n}_m=\hat{a}^\dag_m\hat{a}_m$ is the particle number operator
correspondingly. $J$ describes the hopping strength between adjacent sites which are indicated by the subscript
$\langle m,n\rangle$.
$U$ refers to the on-site interaction between atoms.
$K$ and $\omega$ denote the amplitude and frequency of the time-periodic modulation potential, while
$d$ stands for the constant acceleration force.

The dynamical properties of our system can be obtained by studying
the Heisenberg equations of motion for $\hat{a}_m$. Provided that
the particle number of atoms on each site is large enough, we can
safely adopt the mean-field approximation (MFA) to replace these
operators with their expectation values, {\eg}, $\alpha_m =\langle
\hat{a}_m\rangle/\sqrt{N_L}$, where $N_L$ is the average number of
atoms per site. Then the time-evolution equations for $\alpha_m$ are
expressed as
\begin{eqnarray}
\label{eq:He} i\frac{\partial\alpha_m}{\partial t} &=&
-J(\alpha_{m+1}+\alpha_{m-1})+g\mid\alpha_m\mid
^2\alpha_m\nonumber\\
&&+[K\cos(\omega t)+d]m\alpha_m,
\end{eqnarray}
where the natural unit $\hbar=1$ is taken and $g=N_L U$ denotes the
normalized atomic interaction. Eqs.~(\ref{eq:He}) are also regarded
as discretized Gross-Pitaevskii (GP) equations. Additionally, the
particle number conservation $\sum^{}_m\hat{a}^\dag_m\hat{a}_m=N$
gives $\sum^{}_m|\alpha_m|^2=L$, where $N$ and $L$ are the total
numbers of atoms and optical lattice sites, respectively, and their
ratio is $N_L$, $\ie$, $N_L=N/L$.

In the interaction-free case, \ie, $g=0$, Eqs.~(\ref{eq:He})
describe the single-particle dynamics whose analytical solutions can
be explicitly derived through the gauge transformation~\cite{tilted
the3}
\begin{equation}
\label{eq:Transformation} \alpha_m(t)\rightarrow\exp
\left[-im\left(d t+\frac{K}{\omega}\sin(\omega t)\right)\right]
\tilde{\alpha}_m(t),
\end{equation}
then Eqs.~(\ref{eq:He}) become
\begin{equation}
\label{eq:Heg=0Trans} i\frac{\partial\tilde{\alpha}_m}{\partial
t}=-J\left(e^{-i D(t)}\tilde{\alpha}_{m+1}+e^{i
D(t)}\tilde{\alpha}_{m-1}\right),
\end{equation}
where $D(t)=d t+K/{\omega}\sin(\omega t)$. Because the translation
symmetry is retrieved back in the above equation, we can impose the
spatial periodic boundary conditions, $\ie$,
$\tilde{\alpha}_m(t)=\tilde{\alpha}_{L+m}(t)$. The corresponding
semi-classical Hamiltonian now reads
\begin{equation}
\label{Ham2}
H(t)=-J\sum^{}_m\left(e^{-iD(t)}\tilde{\alpha}^{*}_m\tilde{\alpha}_{m+1}+e^{iD(t)}\tilde{\alpha}^{*}_{m+1}\tilde{\alpha}_m\right).
\end{equation}

The Bloch-wave representation is desirable in order to obtain the solutions for $\tilde{\alpha}_m$,
\begin{equation}
\tilde{\alpha}_m=L^{-1/2} \sum^{}_k e^{i k m}b_k,
\end{equation}
where $k=2\pi n/L$ is the quasimomentum ($-\pi \leq k < \pi$),
$n=0,\pm 1,\ldots,\pm (L-1)/2$ for odd $L$,
while $n=0,\pm 1,\ldots,\pm L/2$ for even $L$.
Then the evolution equations of $b_k$ take the following simple form
\begin{equation}
i\frac{\partial b_k}{\partial t} =
-2J\cos\left[k-dt-\frac{K}{\omega}\sin(\omega t)\right]b_k,
\end{equation}
whose solutions can be explicitly expressed as
\begin{equation}
b_k(t)=b_k(0)\exp \left\{\displaystyle
i2J\left[\cos(k)S(t)-\sin(k)C(t)\right]\right\}.
\end{equation}
Here the factor $b_k(0)$ is the integral constant, and $S(t)$ and $C(t)$
are defined as
\begin{eqnarray}
S(t)&=&\sum^{+\infty}_{n=-\infty}\frac{\sin(n\omega+d)t}{n\omega+d}\mathcal {J}_n\left(\frac{K}{\omega}\right),\label{eq:S} \\
C(t)&=&\sum^{+\infty}_{n=-\infty}\frac{\cos(n\omega+d)t-1}{n\omega+d}\mathcal
{J}_n\left(\frac{K}{\omega}\right)\label{eq:C},
\end{eqnarray}
where $\mathcal {J}_n$ denotes the $n$th ordinary Bessel function.

The analytical solutions $\alpha_m$ for $g=0$ can be directly
obtained through the inverse Fourier transformation. They are the
starting points to derive the trial solutions for $g\neq 0$ case.
Specifically, we are concerned about $b_k(0)=\delta_{k,p}$ case,
which can simplify the final expressions of solutions of
Eqs.~(\ref{eq:He}) as
\begin{eqnarray}
\alpha_m(t)&=&\exp\{\displaystyle
im[p-dt-\frac{K}{\omega}\sin(\omega t)]\}\times
    \nonumber\\
&&\exp\{ \displaystyle i2J[\cos(p)S(t)-\sin(p)C(t)]-i g
t\}.\label{eq:alpha}
\end{eqnarray}
Note that the solutions are applicable for $L \rightarrow \infty$ in which the boundary conditions make no difference.

To explore the dynamical stability of the system, we assume small
fluctuations around the stationary solutions following the usual
ansatz~\cite{periodic the5, tilted the2}, {\ie},
$\alpha_m(t)=\alpha^{0}_m(t)+\delta\alpha_m(t)$, where
$\alpha^{0}_m(t)$ take the expressions as Eqs.~(\ref{eq:alpha}). The
fluctuations can be expressed as
\begin{equation}
\label{eq:p}
\delta\alpha_m(t)=\alpha^0_m(t)[u(t)e^{iqm}+v^*(t)e^{-iqm}].
\end{equation}
Here $q$ is the momentum of the excitation relative to the
condensate.
Substituting the trial solutions into Eqs.~(\ref{eq:He}),
we obtain the Bogoliubov-de Gennes (BdG) equations for the quasiparticle
excitations $u(t)$ and $v(t)$,
\begin{equation}
\label{eq:BdG} i\frac{d}{dt}{u(t)\choose v(t)}=\mathcal
{M}(q,t){u(t)\choose v(t)},
\end{equation}
where the elements of the matrix $\mathcal {M}(q,t)$ are given by
\begin{eqnarray}
\mathcal{M}_{11}(q,t)&=& 4J\sin\left(\frac{q}{2}\right)\sin\left(\frac{q}{2}+p-\frac{K}{\omega}\sin(\omega t)-dt\right)+g,\nonumber\\
\mathcal{M}_{12}(q,t)&=& g=-\mathcal{M}_{21}(q,t),\nonumber\\
\mathcal{M}_{22}(q,t)&=&-4J\sin\left(\frac{q}{2}\right)\sin\left(\frac{q}{2}-p+\frac{K}{\omega}\sin(\omega
t)+dt\right)-g\nonumber.
\end{eqnarray}
Note that the matrix $\mathcal {M}(q,t)$ is also time-periodic with the periodicity
$T$ being the lowest common multiple of $2\pi/\omega$ and $2\pi/d$.

It will be convenient to introduce the evolution operator $U(t)$ in
order to characterize the evolution of $u(t)$ and $v(t)$, $\ie$,
$(u(t), v(t))^T=U(t) (u(0), v(0))^T$. Thus, the dynamical behavior
$U$ is governed by
\begin{equation}
\label{eq:BdG_U} i\frac{d}{dt}U(t)=\mathcal {M}(q,t)U(t).
\end{equation}
Using the $2\times 2$ unit matrix as the initial data, we
numerically solve Eq.~(\ref{eq:BdG_U}) over period $T$. According to
the Floquet theorem, the eigenvalues $\lambda_i$ of $U(T)$
correspond to the excitation quasienergies $\varepsilon_i$ via
$\lambda_i=\exp[-i\varepsilon_i T]$ ($i=1, 2$). The dynamical
stability of the system is specified by the fact that both the
quasienergies $\varepsilon_i$ have no imaginary components, $\ie$,
$|\lambda_i|=1$. If so for all values of $q$, then the solution is
stable. Otherwise, the system may collapse. The analysis of those
features can help us to map out the stability diagrams of the system
in the parameter space of $g/\omega$ and $K/\omega$ which are shown
in Sec.~\ref{sec:dynamical stability}.

\section{Quasienergy}\label{sec:quasienergy}

As stated above, quasienergies are tightly related to the dynamical
stability. Since the period of the time-periodic evolution
matrix in Eqs.~(\ref{eq:Heg=0Trans}) is $T$, the Floquet theorem
enables us to write the solution in the form of $\exp[-i\varepsilon
t]\psi(t)$, where $\varepsilon$ corresponds to the quasienergy of
the system and $\psi(t)$ corresponds to the Floquet state which
shares the same period $T$. In order to obtain the
quasienergies, we pick out all terms which are not $T$-periodic from
$\alpha_m(t)$~\cite{periodic the5}. Note that $g=n\omega+g'$ where $n$ is a positive
integer to ensure $0\leq g' < \omega$. Considering the possible
divergence in the denominators of summations in $S(t)$ and $C(t)$,
we analyze, separately, the cases of $l=d/\omega$ being integer
and non-integer.

For integer $l$, in the limit of $d\rightarrow l\omega$, both
$\exp[\sin(-l\omega+d)t/(-l\omega+d)]\rightarrow t$ and
$\exp[-ig't]$ are not $T$-periodic, then we have the quasienergies
expressed as
\begin{equation}
\varepsilon(p) = -2J\cos(p)\mathcal {J}_{-l}\left(\frac{K}{\omega}\right)+g'.\label{eq:Qe}
\end{equation}
The situation of $d=0$ has already been studied in
Ref.~\cite{periodic the5}, so here we discuss more general case
$d\neq 0$. Throughout our paper, $J$ is set to be the unit. For
example, we show the quasienergy spectrum for $d=\omega$ ($l=1$) and
$d=2\omega$ ($l=2$) in Fig.~\ref{fig:integer}(a) and \ref{fig:integer}(b),
respectively. The spectrum is obtained by the fundamental matrix
method, which requires the diagonalization of
$\vec{\alpha}(0)^T\vec{\alpha}(T)$ where
$\vec{\alpha}(t)=\{\alpha_1(t),\alpha_2(t),\ldots,\alpha_L(t)\}$. As
$L\rightarrow \infty$, the results match the analytical
expressions~(\ref{eq:Qe}) perfectly. Another numerical way to obtain
the quasienergies is to evolve the time-evolution operator of
$\vec{\alpha}(t)$ directly and diagonalize it, which yields the same
results. It is obviously observed from Fig.~\ref{fig:integer}(a) and
\ref{fig:integer}(b) that the spectrum is wholly displaced by the
amount of $g'$ , as stated above. And the collapse points of the
spectrum correspond to the zeroes of $\mathcal {J}_{-l}(K/\omega)$,
which manifests the phenomenon of the ``coherent destruction of
tunneling" (CDT) effect.

For non-integer $l$, the case is much simpler since all terms in
$\alpha_m(t)$ have the periodicity $T$ except $\exp[-ig't]$. One can
conclude that the quasienergies of the system are $\varepsilon(p)=g'$,
independent of the value of $K/\omega$. Our numerical calculations
further confirm our analysis.

\section{DYNAMICAL STABILITY}
\label{sec:dynamical stability}

We are interested in the dynamical stability of the system around the ground state, $\ie$, $p=0$.
To determine the critical interaction $g_c$ beyond which the system becomes unstable, we increase the value of $g$ from zero gradually for each $K/\omega$ until the unstable region is reached.
Then we are able to plot the boundary
between stable and unstable regions in the parameter space of $g/\omega$ and $K/\omega$.

\begin{figure}[tbph]
\includegraphics[width=85mm]{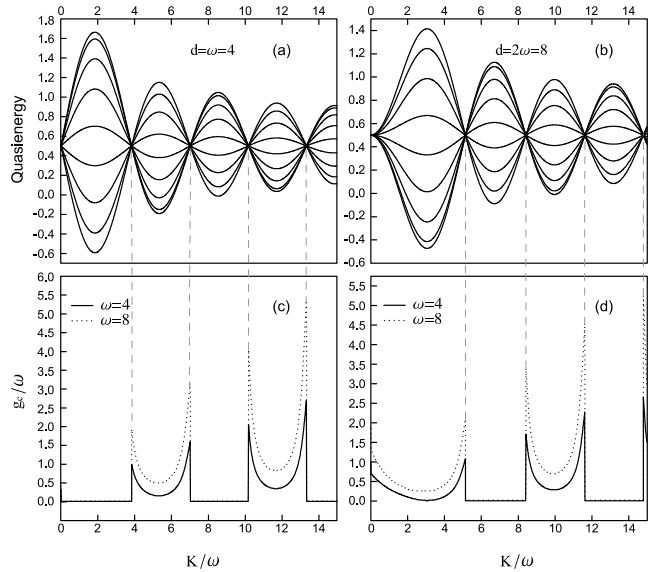}
\caption{Quasienergy spectrum of a nine-site system for (a) $l=1$
and (b) $l=2$ and the $K/\omega$ dependences of the critical
interaction $g_c$ for (c) $l=1$ and (d) $l=2$. The displacement of
quasienergy spectrum is $g=g'=0.5$ for both cases. The vertical
dashed lines are guides to eyes.} \label{fig:integer}
\end{figure}

Since Eq.~(\ref{eq:BdG_U}) can be solved analytically when $g=0$, it
is easy to find that the excitation quasienergies are all real such
that the system is always stable regardless of the value of
$K/\omega$. As $g$ increases, the unstable regions emerge. We
observe different stability diagrams for integer and non-integer
$l$.

For the case of integer $l$, we plot the stability diagrams for
$l=1$ and $l=2$ in Fig.~\ref{fig:integer}(c) and
\ref{fig:integer}(d), respectively. It is clear that the dependence
of $g_{\mathrm{c}}$ on $K/\omega$ shows two kinds of alternative behaviors between plateau and upward concave shapes.
We notice that the separations between the two behaviors appear at the
collapse points of the quasienergy spectrum, also the zeros of
$\mathcal {J}_{-l}(K/\omega)$. From Eq.~(\ref{eq:Qe}) we may define
the effective hopping strength $J_{\mathrm{eff}}(K/\omega)=J\mathcal
{J}_{-l}(K/\omega)$, while the flat critical interaction regions
($g_{\mathrm{c}} \rightarrow 0$) correspond to the negative
$J_{\mathrm{eff}}$. The instability can be understood as the effect
of attractive interactions~\cite{attractive the} for positive
$J_{\mathrm{eff}}$ due to the symmetry of the system
Hamiltonian~\cite{hopping exp}. When $J_{\mathrm{eff}}(K/\omega)=0$,
CDT happens, freezing the dynamics of the system, thus insuring the
stability. $g_{\mathrm{c}}$ peaks around these critical points.

Comparing Fig.~\ref{fig:integer}(c) and \ref{fig:integer}(d) we
notice a seemingly natural, however, interesting phenomenon. When
$l$ is a positive odd integer,  in the limit of infinite small,
however nonzero $K/\omega\rightarrow 0$, the system is always
unstable for any infinite small value of interaction, \ie,
$g_{\mathrm{c}} \rightarrow 0$. The specific choices of $l$ insure
the emergence of the phenomenon which has not been observed in the
case of $l=0$~\cite{periodic the5}.

\begin{figure}[tbph]
\includegraphics[width=60mm]{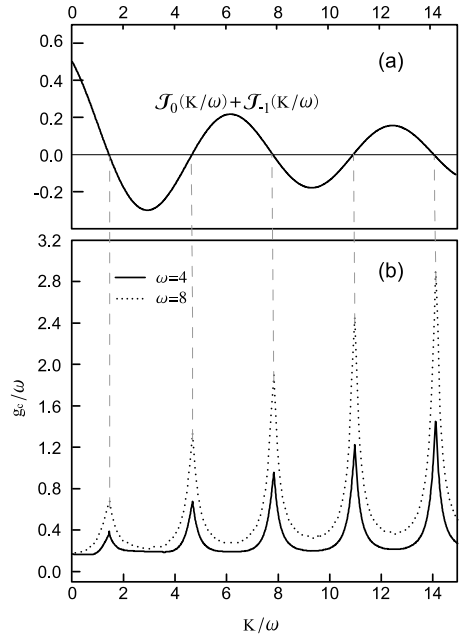}
\caption{The $K/\omega$ dependence of (a) $\mathcal {J}_{0}(K/\omega)+\mathcal {J}_{-1}(K/\omega)$ and (b) the critical interaction $g_{\mathrm{c}}$ for $l=0.5$. The vertical dashed lines are guides to eyes.} \label{fig:non-integer}
\end{figure}

\begin{figure}[tbph]
\includegraphics[width=80mm]{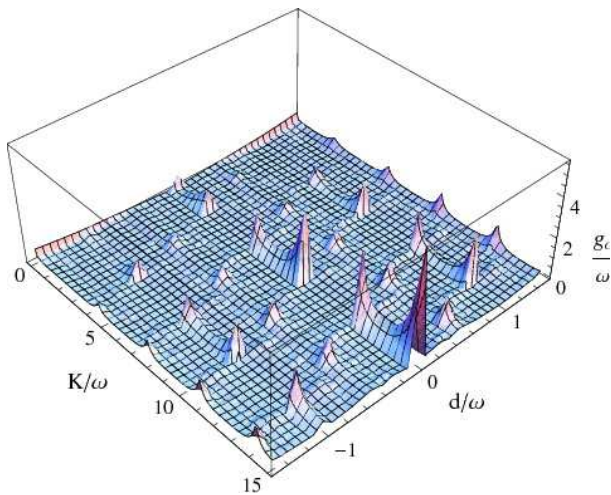}
\caption{(Color online) Dependence of the critical interaction $g_c$ on $K$ and $d$
for $-1.5\leq l\leq 1.5$ in the case of $\omega=4$.}
\label{fig:all}
\end{figure}
For the case of non-integer $l$, the analysis is totally different, of which we start with half-integer $l$ case.
 As an example, we plot the stability
diagram for $l=0.5$ in Fig.~\ref{fig:non-integer}(b). Since the
quasienergies for non-integer $l$ are constant independent of
$K/\omega$, no correspondence between the quasienergies and
$g_{\mathrm{c}}$ peaks is observed, as expected. In order to
determine the positions of $g_{\mathrm{c}}$ peaks approximately, we
have to resort to the expression of $\alpha_m(t)$. Since the
summations appear in the exponent of Eq.~(\ref{eq:alpha}), as
$|n+l|$ increases, the contribution from the $n$-th term reduces
rapidly, if the high modulation frequency limit is assumed, \ie,
$\omega \gg 1/J$~\cite{quasienergy high frequency}. The dominant
terms would be those with the smallest values of $|n+l|$. As an
example, for $l=0.5$, the dominant contributions to $\alpha_m(t)$
would come from $\mathcal {J}_0(K/\omega)+\mathcal
{J}_{-1}(K/\omega)$ which is plotted in
Fig.~\ref{fig:non-integer}(a). The correspondence between the zeros of Bessel functions
and $g_{\mathrm{c}}$ peaks is observed, confirming our analysis. As
stated before, the dynamics of the system is frozen at the zeros of
Bessel functions, indicating the appearance of stable regions, where
$g_{\mathrm{c}}$ peaks with greatest probability. We also notice the
whole displacement of $g_c$ whose value depends on $d$. In
Fig.~\ref{fig:all} we plot the behaviors of dynamical stability
around integer and half-integer $l$, compared with which the
crossover between them shows no explicit dependence of
$g_{\mathrm{c}}$ on the values of $K/\omega$. We may interpret the
behavior as the counterbalance of different Bessel functions. Also
from Fig.~\ref{fig:all}, we can find the same dependence of
$g_{\mathrm{c}}$ on $d$ as found in Ref.~\cite{tilted the3} without
time-periodical modulation, $\ie$, $K/\omega=0$.

Additionally, for both integer and non-integer $l$, we find that the
bigger value of $\omega$ favors the high modulation frequency limit,
also enhances the dynamical stability of the system as shown in
Fig.~\ref{fig:integer}(c), \ref{fig:integer}(d) and
Fig.~\ref{fig:non-integer}(b).

\section{Summary}\label{sec:summary}

We considered the system of BECs in one-dimensional tilted
time-periodically modulated lattices. The ratio $l$ of the acceleration
force $d$ to the modulation frequency $\omega$ is defined to help dividing our
analysis into integer and non-integer resonances. Due to the time-periodicity of the system Hamiltonian, we
gave the explicit expressions of quasienergies for both cases
referring to integer and non-integer resonances by making use of
Floquet theorem. For integer $l$, the quasienergy spectrum shows its
correspondence to the Bessel function $\mathcal {J}_{-l}(K/\omega)$
whose zeros correspond to the collapse points of the spectrum,
indicating the appearance of CDT. However, for non-integer $l$, no
dependence of the quasienergies on $K/\omega$ is found.

We investigated the dynamical stability of the system by
characterizing it in the parameter space of the interaction strength
and the modulation amplitude. For integer $l$, we observed two
alternate behaviors of the dependence of the critical interaction
strength $g_{\mathrm{c}}$ on $K/\omega$: plateau ($g_{\mathrm{c}}
\rightarrow 0$) and upward concave. The complete instability in
horizontally flat regions is brought out by the negative effective
tunneling strength. The separations between the two kinds of regions
correspond to the zeros of Bessel Function $\mathcal
{J}_{-l}(K/\omega)$. We also noticed an interesting phenomenon when
$l$ is a positive odd integer. In such case, the system is always
unstable in the limit of $K/\omega \rightarrow 0$ corresponding to
the infinite small, however, nonzero modulation potential. Among non-integer
resonances, we observed that $g_{\mathrm{c}}$ peaks are centered
around half-integer $l$ for which the completely unstable regions
disappear. Also the whole displacement of $g_{\mathrm{c}}(K/\omega)$
emerges whose value is dependent on $d$. The positions of
$g_{\mathrm{c}}$ peaks in the direction of $K/\omega$ are determined
approximately in the high modulation frequency limit. Compared with
integer and half-integer resonances, the crossovers between them
show no explicit dependence of $g_{\mathrm{c}}$ on the modulation
amplitude. Since dynamical stability is experimentally observable,
these features are expected to be confirmed in experiment.

\section{ACKNOWLEDGMENT}

The work is supported by NSFC Grant No.~10874149 and
partially by PCSIRT Grant No.~IRT0754.

\end{document}